# Electronic, Optical and Elastic Properties of Perovskite CsSnI$_{3-x}$Cl$_x$ (x = 0, 1, 2, 3): Using First Principles Study

Amondulloi S. Burhonzoda [1,*], Dilshod D. Nematov [1,*], Mekhrdod S. Kurboniyon [1, 2], Umar Zafari [1, 3],
Kholmirzo T. Kholmurodov [1, 6, 7], Tomoyuki Yamamoto [4, 5], Farhod Shokir [1]

[1] S.U. Umarov Physical-Technical Institute of the NAST, Dushanbe 734063, Tajikistan
[2] School of Optoelectronic Engineering & CQUPT-BUL Innovation Institute, Chongqing University of
Posts and Telecommunications, Chongqing, 400065, China
[3] Center of Innovative Development of Science and New Technologies, NAST, Dushanbe, 734025, Tajikistan
[4] Faculty of Science and Engineering, Waseda University, Tokyo, 169-8555, Japan
[5] Institute of Condensed-Matter Science, Waseda University, Tokyo, 169-8555, Japan
[6] Joint Institute for Nuclear Research, 141980 Dubna, Russia
[7] Dubna State University, 141980 Dubna, Russia
* Correspondence: dilnem@mail.ru, amondullo.burkhonzoda@mail.ru

**Abstract**. The structural, electronic, optical and elastic properties of perovskites of the α-, β- and γ-phase family CsSnI$_{3-x}$Cl$_x$ (x = 0, 1, 2, 3) are analyzed by ab initio methods using GGA, SCAN and HSE06. It is found that the lattice constants of CsSnI$_{3-x}$Cl$_x$ decrease with decreasing anion size from I to Cl, and when passing from CsSnI$_3$ to CsSnCl$_3$, the band gap width decreases slightly at the beginning and then increases. The optical characteristics were evaluated and it was found that the α-, β-phase of CsSnI$_{3-x}$Cl$_x$ at x=1 has a wide absorption range and high absorption coefficients compared to other Cl concentrations. The elastic constants of α-CsSnI$_{3-x}$Cl$_x$ perovskite have been calculated and their bulk moduli, shear moduli, sound velocities and Debye temperatures under pressure in the range from 0 to 30 GPa have been considered, with the external pressure induction increasing in steps of 5 GPa. These materials are found to be mechanically stable perovskites according to Born stability criteria and low bulk moduli. Thus, we predict that perovskites of the CsSnI$_{3-x}$Cl$_x$ family are good candidates for optical materials used in solar cells.

## 1. Introduction

Due to population growth and the resulting increase in the need for electrical energy, the last decade has seen the search for high-efficiency materials to create and develop solar cells by various methods [1-3]. The most common material capable of absorbing sunlight and converting it into electrical energy is silicon. In principle, silicon is the basis for most modern single junction solar cells. However, silicon does not have an easily tunable band gap and is not suitable for photovoltaic or multi-junction solar cells, and has a number of limitations and low efficiency.

As a result of numerous experimental studies and theoretical calculations, some photovoltaic materials such as copper-indium-gallium selenide (CIGS), cadmium telluride (CdTe) and iron disulfide (FeS$_2$) have been found and studied, which have much higher efficiency compared to silicon [10-16]. In addition, much higher efficiencies can be achieved with multilayer cells made of other semiconductor materials. For example, the efficiency of solar cells made of AlGaInP, AlGaAs, GaAs is 47.1% [25]. However, all these compounds are expensive due to the presence of scarce elements, and their production on an industrial scale cannot be realized yet.

Meanwhile, perovskite crystals have attracted much research attention due to their important chemical, electronic, mechanical and photovoltaic properties such as high absorption capacity, high carrier mobility, low exciton binding energy and attractiveness for use in solar cells [4-9]. Metal halide perovskites having the form ABX$_3$ (here A = Cs, Rb, MA (CH$^{+3}$NH$^{3+}$), FA (CH(NH$^{+2}$)$^{2+}$); B = Pb$^{2+}$, Ge$^{2+}$; X = I$^-$, Br$^-$, Cl$^-$ or mixed X- anions) show great potential for solar cell applications due to their excellent absorption coefficient, long lifetime and easily controlled band gap [17-25]. Among these compounds, the use of lead halide perovskites (CsPbX$_3$) as absorbers in solar cells has several advantages and disadvantages over other halide perovskites. Their small band gap and good absorption capacity make them a good material for optoelectronic applications, but these structures are unstable at ambient temperature, which hinders their practical use [28-29] and commercialization [30]. However, one of the major disadvantages of Pb is that Pb$^{2+}$ is toxic and Pb-based perovskites pose serious health and environmental hazards. Despite the breakthrough of 26.1% (USTC) energy conversion efficiency demonstrated in organic-inorganic hybrid perovskite solar cells [26-27], there are still critical issues related to instability and toxicity that could potentially hinder their commercialization.

Therefore, replacement of lead by other non-toxic elements is highly desirable. Recent review articles [35-37] emphasized the need to develop new perovskite materials with a direct band gap and outlined the scientific problems associated with lead-free perovskites. Thus, to obtain lead-free material for photovoltaic applications, the B site in ABX$_3$ (B = Pb$^{2+}$) can be replaced by other divalent metals such as Sr$^{2+}$, Ge$^{2+}$, Zn$^{2+}$, Sn$^{2+}$ [31-34]. The most suitable substitution for the Pb position is the divalent metal Sn because Sn and Pb belong to group 14 of Mendeleev's periodic table and hence have similar physicochemical properties [38-40].

Understanding the structural, electron-optical and mechanical properties of these materials for use in α-, β- and γ-phase CsSnI$_{3-x}$Cl$_x$ hybrid solar cells is highly desirable, which is investigated in this work.

## 2. Computational methods

The calculations were performed using the Vienna ab initio simulation package (VASP) [41] in the framework of density functional theory (DFT). At the first stage of the study, a thorough convergence check on the cutoff and KPOINS energies was carried out, resulting in their optimal values for further calculations.

Thus, a value of 500 eV was chosen as the cutoff energy for all CsSnI$^{3-x}$Cl$^x$ phases. The Monkhorst-Pack k-point grid sampling was set to $4 \times 4 \times 4$, $5 \times 5 \times 7$, and $3 \times 3 \times 2$ for α- (cubic Pm-3m (221)), β- (tetragonal P4/mbm (127)), and γ- (rhombic Pnam (62)) phases, respectively. All crystals were optimized using the same software and under the same conditions. The generalized gradient approximation proposed by Purdue, Burke, and Ernzerhof (GGA-PBE) [42] was used to express the electron-electron correlation, while the more accurate meta-GGA (SCAN) [43] and hybrid (HSE06) [44] functionals were used to calculate the electronic properties, since GGA usually underestimates the band gap magnitude for these compounds.

The elastic properties were calculated under the same conditions using the GGA-PBE approximation. The k-point value of the Monkhorst-Pack k-point grid was doubled to calculate the DOS. The optical properties of CsSnI$_{3-x}$Cl$_x$ were calculated from the SCAN-optimized structure using the GGA [57] -mBJ [58] potencies implemented in the WIEN2K package [59] with 10×10×10 k-point Brillouin zone.

Machine learning models [72] can give a general idea of the expected value of the bandgap for inorganic and halide perovskites. A linear regression model using Python code was used to predict the band gap of α-, β-, and γ-phases of CsSnI$_{3-x}$Cl$_x$ [62].

## 3. Results and discussion

### 3.1. Structural properties

CsSnI$_{3-x}$Cl$_x$ (x = 0, 1, 2, 3) perovskites exist in different phases and in the present studies we focus only on α-, β-, γ- phases using DFT method, which respectively contain 5, 10, 20 atoms in the elemental cell. Figure 1(a), (b), (c) shows the optimized crystal structures of α-, β-, γ- phases of CsSnI$_3$, respectively.

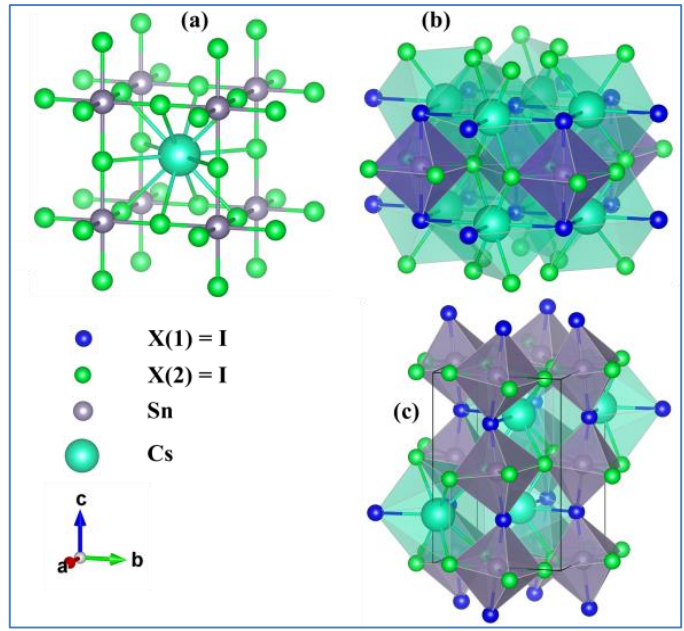

Figure 1. Crystal structures of (a) α-, (b) β-, (c) γ- phases of CsSnI$_3$

As can be seen from Figure 1, the β-, γ- phases of CsSnI$_3$ have 2 types of atom I, and site occupation was carried out to obtain the mixed structures. It is found that the I(1) position is more energetically favorable for Cl ions in CsSnI$_3$.

Thus, the I atom in the α-, β-, γ- phases of the CsSnI$_3$ structure was replaced by Cl atoms, which were then optimized perovskite structures CsSnI$_2$Cl CsSnICl$_2$ and CsSnCl$_3$. The obtained lattice constants are shown in Table 1 and the obtained values were in a great agreement with other theoretical and experimental values. We find that the lattice constants of CsSnI$_{3-x}$Cl$_x$ decrease with decreasing the size of anions from I to Cl.

For optoelectronic materials, it is important to guarantee the structural and thermodynamic stability of perovskite CsSnI$_{3-x}$Cl$_x$ compounds, which can be evaluated using the Goldschmidt tolerance factor and octahedral factor, according to equations (1) and (2), respectively [60-61].

$$t_G = \frac{R_A + R_X}{\sqrt{2}\,(R_B + R_X)},\tag{1}$$

$$\mu = \frac{R_B}{R_A},\tag{2}$$

where $R_A$ and $R_B$ are the effective radius of Cs$^+$ and Sn$^{2+}$, respectively, $R_X$ is the effective radius of the I$^-$, Cl$^-$ and their average value.

The stable three-dimensional structure of halide perovskite is usually formed at $0.80 \leq t_G \leq 1.0$ and $0.44 \leq \mu \leq 0.9$. For CsSnI3-xClx, $\mu = 0.61$ and $t_G$ values are in the range of (0.849 - 0.869), indicating a stable perovskite structure.

Table 1. Calculated lattice parameters of CsSnI$_{3-x}$Cl$_x$ (x = 0, 1, 2, 3) in Å

| System | | This work | | Other works |
|---|---|---|---|---|
| | | GGA | SCAN | |
| CsSnI$_3$ | α (a=b=c) | 6.261 | 6.179 | 6.22 [45,46]; 6.231 [49]; 6.143 [50]; 6.219 [51]; 6.28 [54]; 6.22 [56] |
| | β (a=b,c) | 8.789, 6.318 | 8.680, 6.239 | 8.772, 6.261 [45]; |
| | γ (a,b,c) | 8.957, 8.667, 12.503 | 8.847, 8.533, 12.372 | 8.688, 8.643, 12.378 [45]; |
| CsSnI$_2$Cl | α (a=b=c) | 6.306 | 6.208 | - |
| | β (a=b,c) | 8.367, 6.319 | 8.270, 6.225 | - |
| | γ (a,b,c) | 8.327, 8.327, 12.592 | 8.220, 8.220, 12.394 | - |
| CsSnICl$_2$ | α (a=b=c) | 6.291 | 6.167 | - |
| | β (a=b,c) | 8.467, 5.631 | 7.779, 6.195 | - |
| | γ (a,b,c) | 7.884, 7.884, 12.604 | 7.771, 7.772, 12.385 | - |
| CsSnCl$_3$ | α (a=b=c) | 5.617 | 5.538 | 5.56 [46]; 5.604 [52]; 5.618, 5.615 [53]; 5.60 [55]; 5.61 [54]; |
| | β (a=b,c) | 7.898, 5.654 | 7.792, 5.583 | - |
| | γ (a,b,c) | 7.956, 7.884, 11.236 | 7.840, 7.778, 11.088 | - |

## 3.2. Electronic properties

The electronic properties of the CsSnI$_{3-x}$Cl$_x$ are calculated by using GGA, SCAN and HSE06. The HSE06 functional is the most appropriate for the evaluation of the band structure and estimate band gaps comparable with the experiment. So, this functional is highly reliable for studying the electronic properties of perovskite-like systems, in comparison with GGA, which greatly underestimates the band gap. Therefore, to understand the electronic conductivity of the CsSnI$_{3-x}$Cl$_x$ system (x = 0, 1, 2, 3), this paper considers the band gap magnitude, total density of states (TDOS) and partial density of states (PDOS) obtained by calculating the hybrid functional HSE06.

Table 2 summarizes the calculated band gap values and compares them with experimental and other theoretical data. The good agreement of the results for pure perovskites confirms the predicted values of the band gap of other phases of pure and mixed systems. The variation of the band gap width of α-, β- and γ- CsSnI$_{3-x}$Cl$_x$ (x = 0, 1, 2, 3) is presented in Figure 2, where it can be seen that the band gap first slightly decreases and then increases when going from CsSnI$_3$ to CsSnCl$_3$. This trend in all investigated CsSnI$_{3-x}$Cl$_x$ phases propagates in a similar way. This is due to the fact that the I atom has a larger size than the Cl atom. The gradual substitution of the I atom with Cl atoms leads to a decrease in the atom size and further increases the separation between the conduction band (CB) and valence band (VB).

To evaluate the material in terms of photovoltaic applications, the value of the band gap should be in the range from 1.0 to 1.5 eV. As suggested by Shockley and Kweisser, the maximum efficiency in this interval can be achieved at a band gap width of 1.34 eV [63-64]. From this point of view, according to calculations based on the hybrid potential of HSE06, all investigated structures are good and potential materials for photovoltaic applications, except for CsSnCl₃, whose band gap exceeds the effective limit.

Table 2. Calculated and experimental band gap
of α-, β- and γ- CsSnI$_{3-x}$Cl$_x$ (x = 0, 1, 2, 3) in eV

| System | | This work | | | | Exp. |
|---|---|---|---|---|---|---|
| | | GGA | SCAN | HSE06 | ML | |
| CsSnI3 | α | 1.115 | 1.122 | 1.326 | 1.232 | - |
| | β | 0.942 | 1.108 | 1.230 | 0.876 | - |
| | γ | 1.161 | 1.146 | 1.435 | 1.171 | 1.3[48] |
| CsSnI₂Cl | α | 1.093 | 1.074 | 1.331 | 1.391 | - |
| | β | 0.592 | 0.538 | 0.834 | 1.063 | - |
| | γ | 0.791 | 0.708 | 1.036 | 1.294 | - |
| CsSnICl₂ | α | 1.202 | 1.207 | 1.482 | 1.564 | - |
| | β | 0.658 | 0.581 | 0.919 | 1.156 | - |
| | γ | 1.062 | 0.985 | 1.379 | 1.511 | - |
| CsSnCl₃ | α | 1.520 | 1.585 | 1.845 | 1.745 | 1.9[47] |
| | β | 1.359 | 1.472 | 1.758 | 1.360 | - |
| | γ | 1.646 | 1.718 | 2.026 | 1.707 | - |

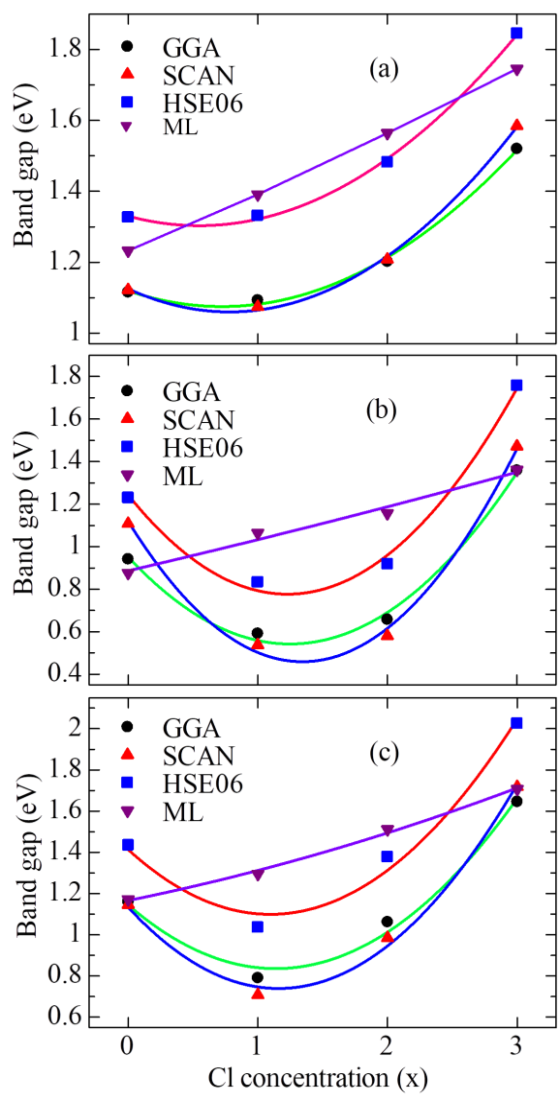

Figure 2. Calculated band gap of (a)- α, (b)- β and (c)- γ phases CsSnI$_{3-x}$Cl$_x$ as function of Cl concentration x (0, 1, 2, 3), with GGA, SCAN, HSE06 functionals and ML

The total and partial densities of states of α-, β-, and γ-phases of CsSnI$_{3-x}$Cl$_x$ (x = 0, 1, 2, 3) are shown in Figures 3 and 4, respectively. The valence band maximum is zero and the dashed lines indicate the Fermi level. It can be seen from the presented figures that none of the studied structures crosses the Fermi level, and this confirms that these materials exhibit semiconducting properties. In the valence band and conduction band, the γ-phase of CsSnI$_{3-x}$Cl$_x$ has a large number of energy levels in the unit energy range, followed by β- and α-phases, respectively.

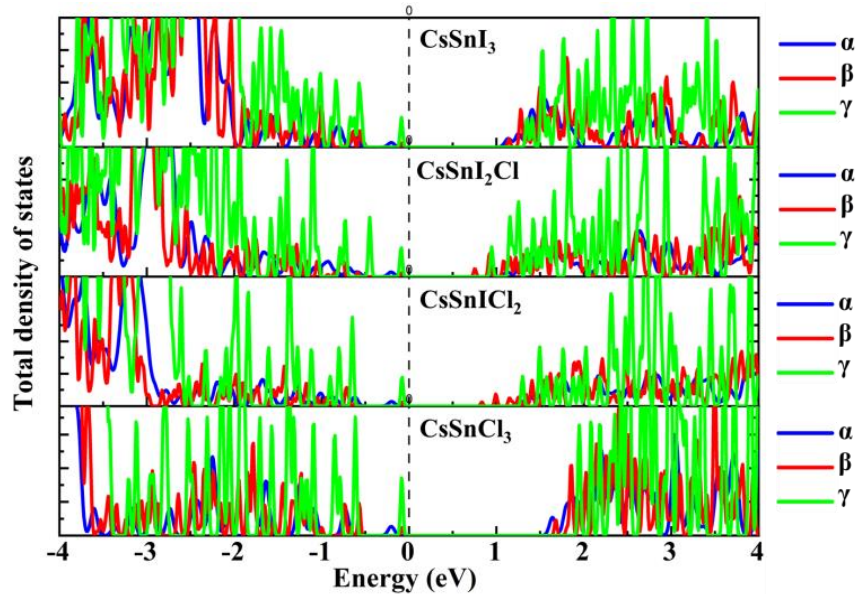

Figure 3. Total densities of states of α-, β-, and γ-phases
of CsSnI3-xClx calculated from the HSE06 hybrid functional

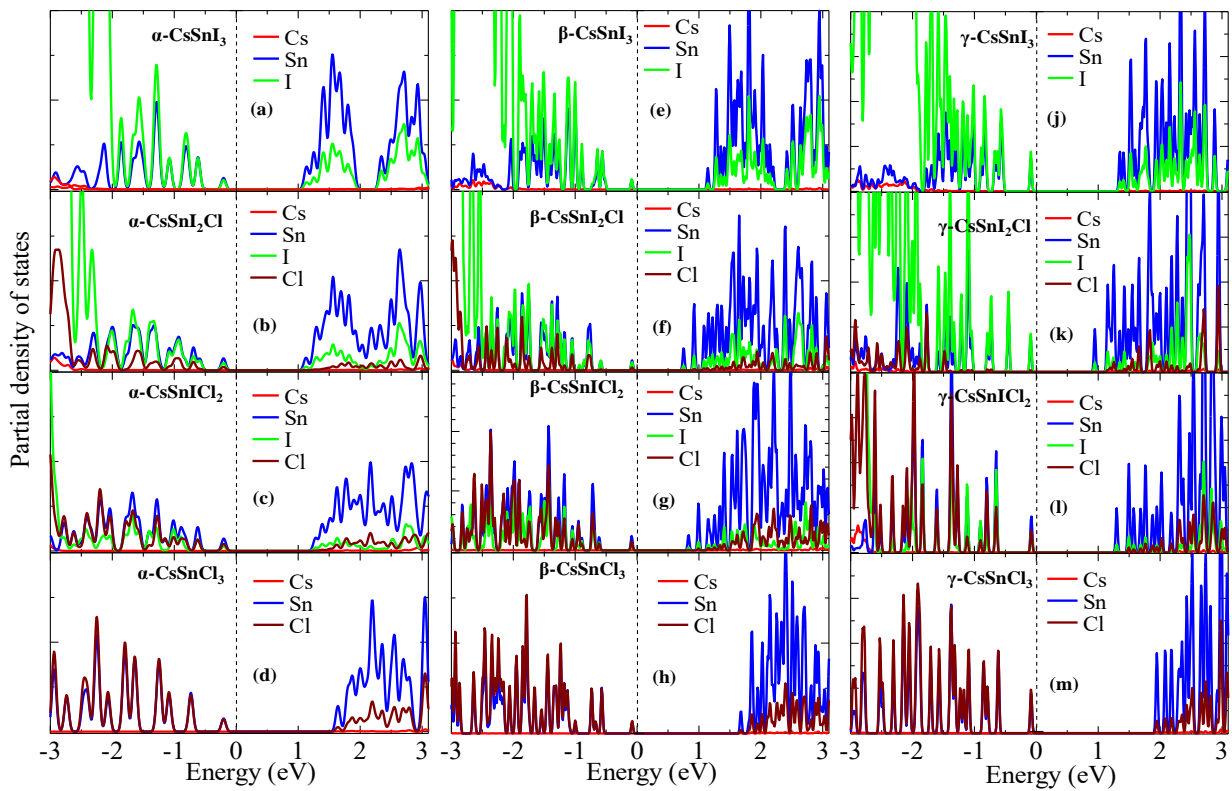

Figure 4. Partial densities of state of α-, β- and γ-phases of
CsSnI3-xClx calculated from the hybrid functional HSE06

The partial densities of electronic states (Figure 4) show the contribution of individual atoms of α-, β-, and γ-phases of CsSnI$_{3-x}$Cl$_x$ (x = 0, 1, 2, 3). In all systems, the contribution of electrons of the Cs atom is very little noticeable in the formation of the valence band (left part of the Fermi level) and conduction band (right part of the Fermi level). It can be said that the Cs atom practically does not participate in the formation of energy bands. The PDOS picture shows that the main contribution to the formation of the valence band and conduction band is made by Sn atoms and halogens, which play an important role in the appearance of the band gap. In the formation of the valence band of the α-, β- and γ-phases of CsSnI$_3$, the main contribution is made by the electrons of atom I, and the contribution of the electrons of atom Sn is much smaller than that of atom I. In the conduction band formation, on the contrary, the contribution of Sn atom electrons is larger than that of I atom electrons. This tendency similarly extends to the other investigated systems. In the transition from I to Cl in the formation of the valence band and conduction band, the contribution of electrons of Cl atoms increases and plays a major role in the increase of the band gap. In contrast to other systems, the electrons of Sn and Cl atoms have almost equal shares in the formation of the valence band of α-, β-, and γ-phases of CsSnCl$_3$.

### 3.3. Optical properties

Optical functions and optical characteristics of materials are crucial parameters to better understand the electronic configuration of materials and to determine their suitability for photovoltaic applications, since they provide important information about its interaction with light. Therefore, the study of optical properties of these halide perovskites is very important and their complex dielectric function $\varepsilon(\omega) = \varepsilon_1(\omega) + i\varepsilon_2(\omega)$ plays a fundamental role in all optical property parameters. Here $\varepsilon_1(\omega)$ and $\varepsilon_1(\omega)$ represent the real and imaginary parts of the dielectric function, respectively. Using $\varepsilon_1(\omega)$ and $\varepsilon_1(\omega)$ other parameters such as absorption coefficient $\alpha(\omega)$, reflectivity $R(\omega)$, refractive index $n(\omega)$, extinction coefficient $k(\omega)$, optical conductivity $\sigma(\omega)$ and energy loss function $L(\omega)$ can be obtained [65-68]:

$$\alpha(\omega) = \sqrt{2\omega}\left[\sqrt{\varepsilon_1^2(\omega) + \varepsilon_2^2(\omega)} - \varepsilon_1(\omega)\right]^{1/2}, \qquad (3)$$

$$R(\omega) = \left|\frac{\sqrt{\varepsilon_1(\omega) + i\varepsilon_2(\omega)} - 1}{\sqrt{\varepsilon_1(\omega) + i\varepsilon_2(\omega)} + 1}\right|^2 = \frac{(n-1)^2 + k^2}{(n+1)^2 + k^2}, \qquad (4)$$

$$n(\omega) = \left[\sqrt{\varepsilon_1^2(\omega) + \varepsilon_2^2(\omega)} + \varepsilon_1(\omega)\right]^{1/2}/\sqrt{2}, \qquad (5)$$

$$L(\omega) = \frac{\varepsilon_2(\omega)}{\varepsilon_1^2(\omega) + \varepsilon_2^2(\omega)}, \qquad (6)$$

$$k(\omega) = \left[\frac{\sqrt{\varepsilon_1^2(\omega) + \varepsilon_2^2(\omega)} - \varepsilon_1}{2}\right]^{1/2}, \qquad (7)$$

$$\sigma(\omega) = \frac{i(\omega)}{4\pi}\varepsilon(\omega). \qquad (8)$$

The calculated real and imaginary parts of the dielectric functions of α-, β- and γ-phases of CsSnI$_{3-x}$Cl$_x$ (x = 0, 1, 2, 3) are plotted in the range of 0-14 eV (Figure 5). $\varepsilon_1(\omega)$ represents the stored energy of the medium or material that can be given up. $\varepsilon_2(\omega)$ explains the absorption capacity and behavior of these materials for photovoltaics. As can be seen from Figures 5(a),(b),(c), the static permeability $\varepsilon_1(0)$ is determined by the lower energy limit $\varepsilon_1(\omega)$. The $\varepsilon_1(\omega)$ peaks shift toward higher energies in the transition from I to Cl. The calculated values of $\varepsilon_1(0)$ for α-, β- and γ-phases of CsSnI$_{3-x}$Cl$_x$ (x = 0, 1, 2, 3) using the GGA-mBJ potential are summarized in Table 3. $\varepsilon_1(0)$ and the band gap are in inverse dependence on each other. Figure 9 (a) shows that $\varepsilon_1(\omega)$ for the cubic phase of $CsSnI_3$, $CsSnI_2Cl$, $CsSnICl_2$ and $CsSnCl_3$ increases and reaches its maximum value of 7.71, 8.24, 4.49 and 4.65 at 1.4, 0.8, 1.51 and 1.64 eV, respectively. Figure 9 (b) shows that $\varepsilon_1(\omega)$ for the tetragonal phase of $CsSnI_3$, $CsSnI_2Cl$, $CsSnICl_2$ and $CsSnCl_3$ increases and reaches its maximum value of 7.27, 8.13, 4.69 and 4.41 at 1.80, 0.73, 1.5 and 1.91 eV, respectively. Figure 9 (c) shows that $\varepsilon_1(\omega)$ for the orthorhombic phase of $CsSnI_3$, $CsSnI_2Cl$, $CsSnICl_2$ and $CsSnCl_3$ increases and reaches its maximum value of 7.38, 5.89, 4.30 and 4.49 at 1.83, 1.67, 1.8 and 1.89 eV, respectively. Then at higher energy values, $\varepsilon_1(\omega)$ becomes negative, which is explained by the fact that in this range, $CsSnI_{3-x}Cl_x$ compounds will behave as a metal with high reflectivity of incident light. Based on the graphs and the given energy values at which the first peaks appear, it can be stated that these compounds exhibit good electro polarization properties in the visible range.

The imaginary parts of the dielectric function of $CsSnI_{3-x}Cl_x$ for all x concentrations are shown in Figures 5(d), (e), (f) for α-, β-, γ- phases, respectively. It can be seen that the threshold values of $\varepsilon_2(\omega)$ appear at 0.7 eV to 1.7 eV in general for the α-, β-, and γ- phases of $CsSnI_{3-x}Cl_x$ (x = 0, 1, 2, 3) , which is closely related to the band gap values. The first maximum peaks $\varepsilon_2(\omega)$ for α-, β- and γ- phases of $CsSnI_{3-x}Cl_x$ (x = 0, 1, 2, 3) occur at about 2.5 to 4 eV and then tends to 0 at high energies.

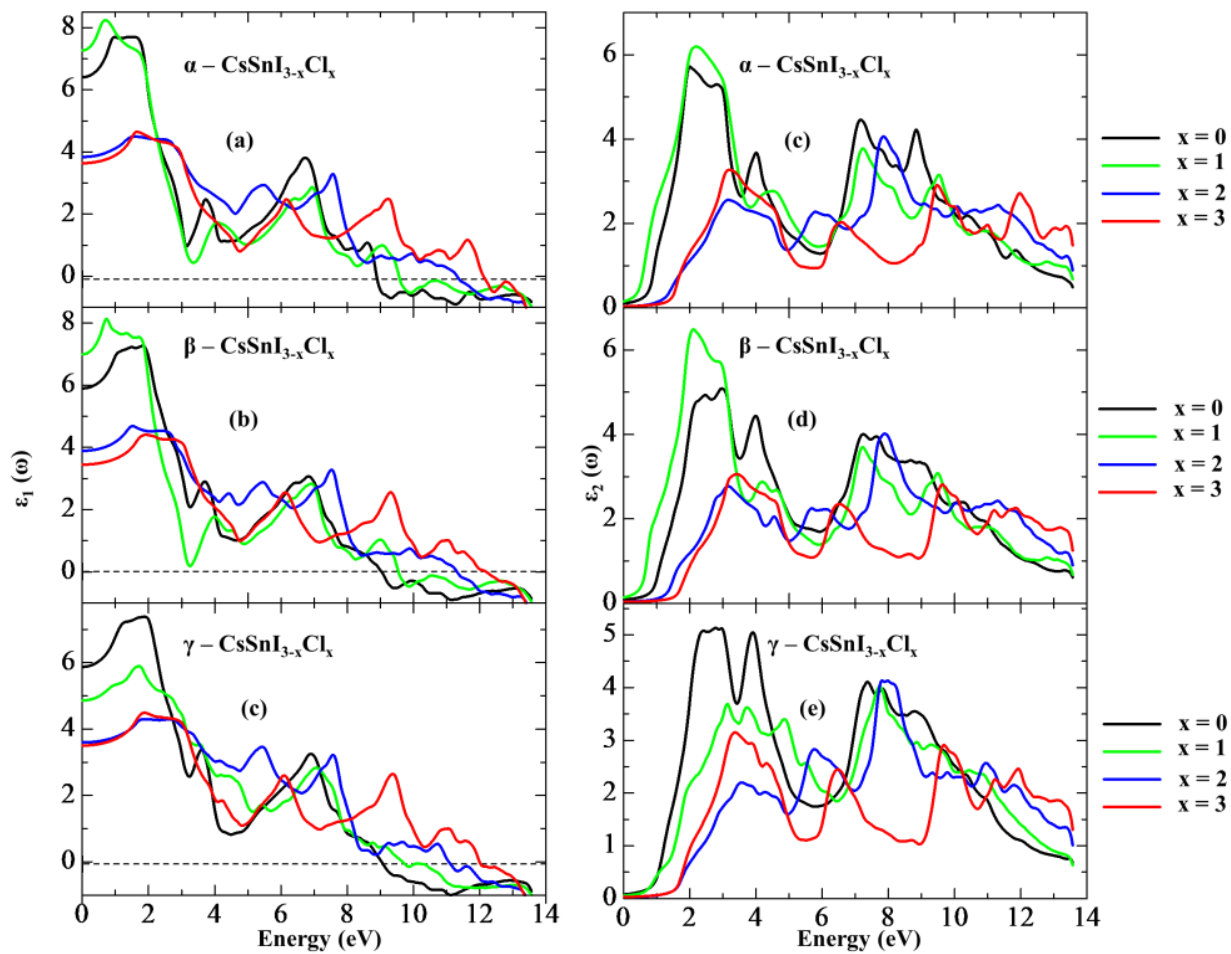

Figure 5. Calculated real part $\varepsilon_1(\omega)$ and imaginary part $\varepsilon_2(\omega)$ of dielectric permittivity as a function of energy using GGA - mBJ potential for α-, β- and γ- $CsSnI_{3-x}Cl_x$ (x = 0, 1, 2, 3)

The refractive index $n(\omega)$ and reflectivity $R(\omega)$ were calculated for α-, β-, and γ-phases of $CsSnI_{3-x}Cl_x$ (x = 0, 1, 2, 3) and are shown in Figure 6. The spectrum of $n(\omega)$ shown in Figure 6 (a), (b), (c) is very similar to the spectrum of $\varepsilon_1(\omega)$. The value of $n(\omega)$ describes the transparency and dispersion characteristics.

Table 3. Calculated and experimental band gap of α-, β- and γ- CsSnI$_{3-x}$Cl$_x$ (x = 0, 1, 2, 3) in eV.

| System | Phases | This work | | |
|---|---|---|---|---|
| | | $\varepsilon_1(0)$ | $n(0)$ | $R(0)$ |
| CsSnI3 | α | 6.41 | 2.53 | 0.188 |
| | β | 5.90 | 2.43 | 0.174 |
| | γ | 5.88 | 2.42 | 0.173 |
| CsSnI$_2$Cl | α | 7.27 | 2.70 | 0.211 |
| | β | 7.00 | 2.60 | 0.204 |
| | γ | 4.86 | 2.21 | 0.141 |
| CsSnICl$_2$ | α | 3.84 | 1.96 | 0.101 |
| | β | 3.89 | 1.97 | 0.107 |
| | γ | 3.60 | 1.90 | 0.095 |
| CsSnCl$_3$ | α | 3.64 | 1.91 | 0.097 |
| | β | 3.45 | 1.86 | 0.090 |
| | γ | 3.51 | 1.87 | 0.092 |

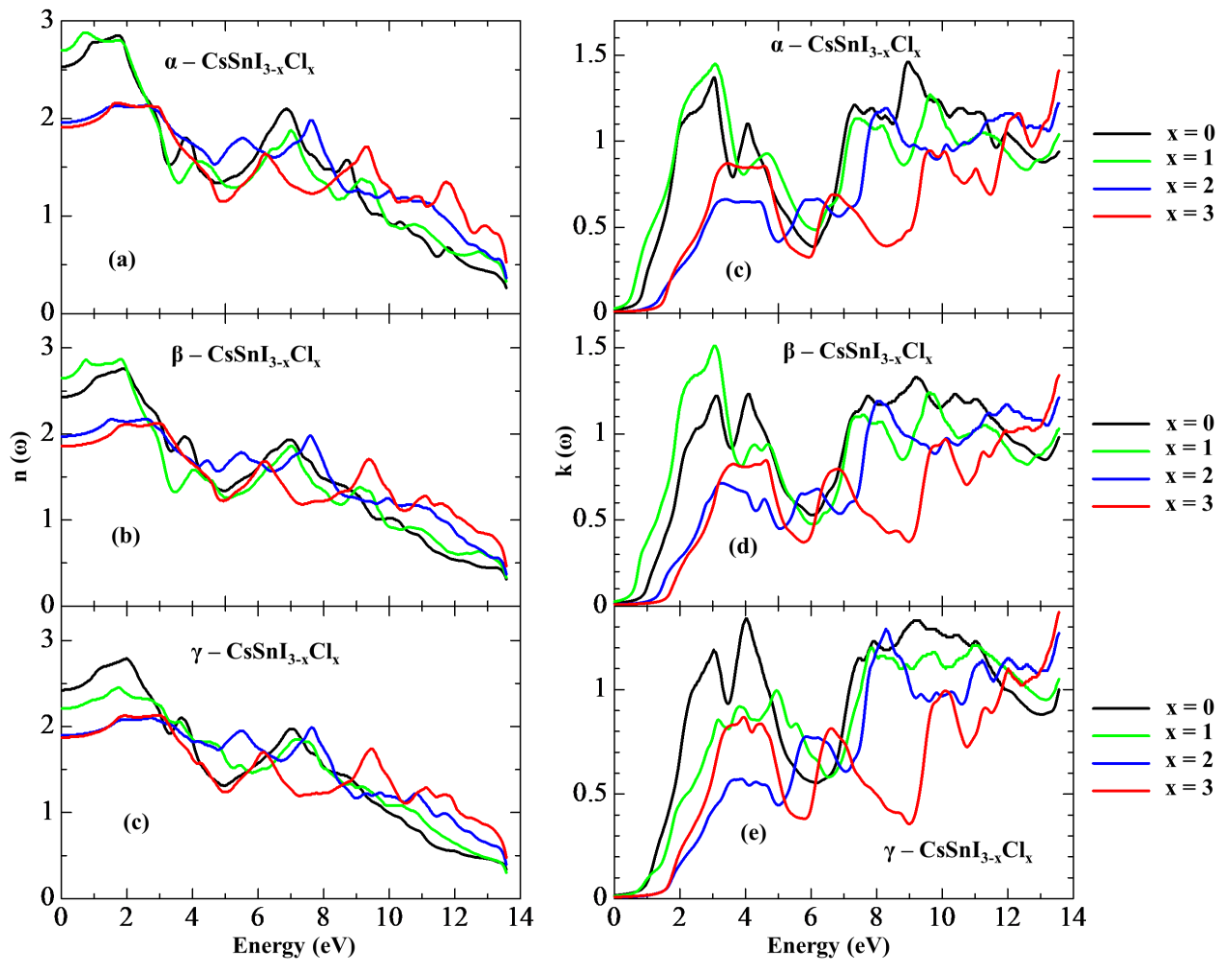

Figure 6. Calculated refractive index n(ω), коэффициент экстинкции $k(\omega)$ spectra of CsSnI$_{3-x}$Cl$_x$ perovskite using the GGA-mBJ potential

The calculated $n(0)$ for α-, β- and γ-phases of CsSnI$_{3-x}$Cl$_x$ (x = 0, 1, 2, 3) using the GGA-mBJ potential are presented in Table 3, whose value lies in the range of 1.8-2.7, which is desirable for optoelectronic applications. For the α- and β-phases of CsSnI$_{3-x}$Cl$_x$, the highest value is achieved at x=1 than other concentrations, while for the γ-phase, the highest value is achieved at concentration x = 0. The first peaks in all systems are reached within 2 eV, then start to decrease and tend to 0 at high energies. This indicates that the refractive index of these perovskites is maximal at low photon energies.

The extinction coefficient $k(\omega)$ shown in Figure 6(d), (e), (f), demonstrates energy absorption and exhibits properties similar to those of $\varepsilon_2(\omega)$. The extinction coefficient ($k$) approaches a maximum value corresponding to the absorption maximum and then decreases as the absorption in the material decreases. The extinction coefficient for α-, β- and γ-phases of CsSnI$_{3-x}$Cl$_x$ (x = 0, 1, 2, 3) is shifted towards lower energies, with the first sharp peak appearing at an energy of about 3 eV and the second at an energy of about 9 eV, indicating better extinction coefficients ($k$) for these materials. However, high extinction coefficients ($k$) are observed for α-, β-phases of CsSnI$_{3-x}$Cl$_x$ at x = 1 compared to other compounds.

Fig. 7(a), (b), (c) shows the energy dependence of the absorption coefficient $\alpha(\omega)$ of α-, β-, and γ-phases of CsSnI$_{3-x}$Cl$_x$ (x = 0, 1, 2, 3). The absorption edge starts from energy 1 - 1.7 eV, which corresponds to the band gap of these materials. CsSnI$_3$ and CsSnI$_2$Cl have good absorption ability than CsSnICl$_2$ and CsSnCl$_3$ due to the smaller band gap in all investigated CsSnI$_{3-x}$Cl$_x$ phases. Due to the large band gap of α-, β- and γ-phases, CsSnCl$_3$ has weak light absorption in the visible region compared to other compounds, but their light absorption is significant in the UV region as well. It can be seen that the first characteristic peaks of the α-phases CsSnI$_3$ and CsSnI$_2$Cl appear in the vicinity of 3 eV, and for the α-phases CsSnICl$_2$ and CsSnCl$_3$ in the vicinity of 4.5 eV.

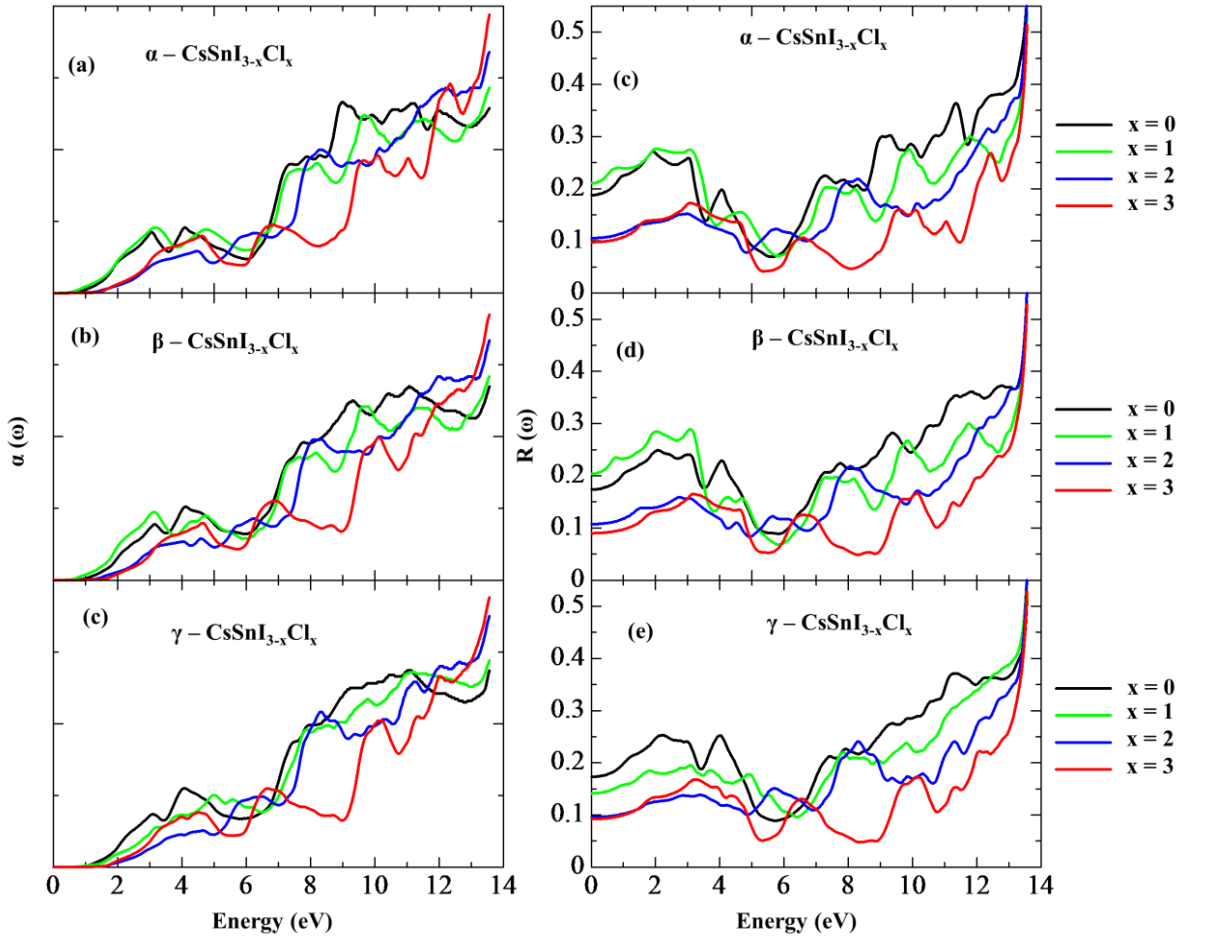

Figure 7. Calculated absorption coefficients spectra α(ω) and reflectivity $R(\omega)$ of CsSnI$_{3-x}$Cl$_x$ perovskite using the GGA-mBJ potential

A similar trend is observed for the other CsSnI$_{3-x}$Cl$_x$ phases, but depending on the band gap, the appearance of the characteristic first peaks is observed at slightly higher energy values. In general, at x=1, an improvement in the absorption coefficient is observed, which is superior to the other studied compounds. Materials used in optoelectronic applications should have good extinction in the visible and near-UV range. Based on the results obtained such as absorption coefficient, $\alpha(\omega)$, it can be stated that CsSnI$_{3-x}$Cl$_x$ is a good material for optoelectronic applications.

The reflectivity $R(\omega)$ determining the nature of the surface of materials for optoelectronics and solar cells are presented in Figure 7(d), (e), (f) for α-, β- and γ- phases of CsSnI$_{3-x}$Cl$_x$ (x = 0, 1, 2, 3), respectively. The calculated values of $R(0)$ for the investigated systems are presented in Table 3. From the results obtained, it can be seen that the reflectivity is very low for all the studied perovskite compounds. However, in α- and β- phases, the reflectivity shows a maximum at x = 1 and a minimum at x = 3, in the optical range. In γ- phases of CsSnI$_{3-x}$Cl$_x$, the reflectivity shows a maximum at x = 0. $R(\omega)$ finds a first maximum peak in the vicinity from 4 eV and minimum values in the vicinity of 6 eV for all the studied systems, then starts to increase with increasing photon energies.

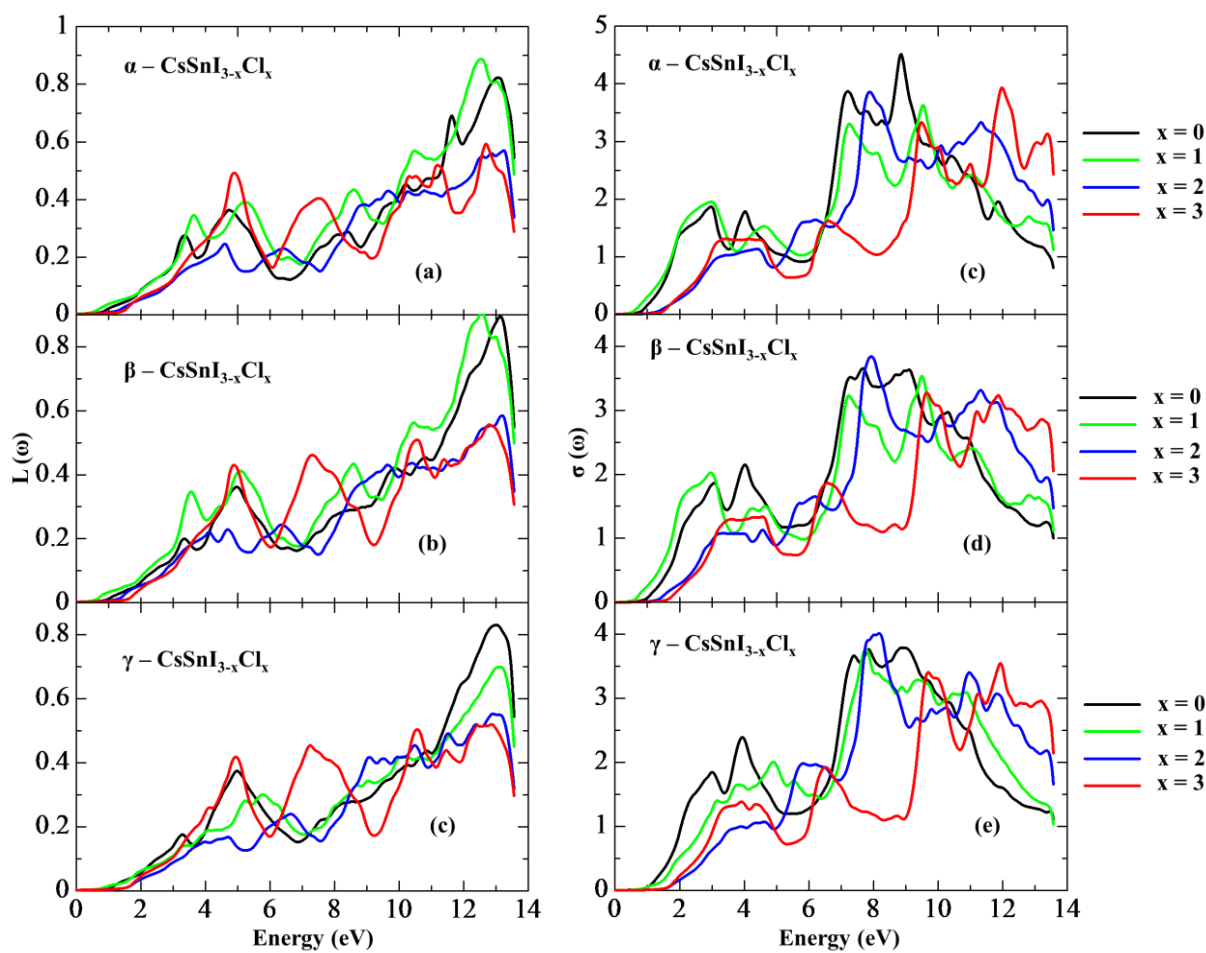

Figure 8. Calculated функция потерь энергии L(ω) и оптической проводимости $\sigma(\omega)$ of CsSnI$_{3-x}$Cl$_x$ perovskite using the GGA-mBJ potential

The electron energy loss function $L(\omega)$, which measures the energy loss during propagation inside the material, is shown in Figure 8 (a), (b), (c) for the α-, β-, and γ-phases of CsSnI$_{3-x}$Cl$_x$ (x = 0, 1, 2, 3), respectively. The peaks in the plots of the electron energy loss function for CsSnI$_{3-x}$Cl$_x$ characterize that the energy loss occurs when the incident photon energy is above the band gap of the material. $L(\omega)$ has a small value up to 10 eV and increases above it for all investigated systems.

The optical conductivity $\sigma(\omega)$ is another important parameter that determines the photovoltaic applications of materials. Figure 8 (d), (e), (f) show the calculated spectra of $\sigma(\omega)$ for α-, β- and γ-phases of CsSnI$_{3-x}$Cl$_x$ (x = 0, 1, 2, 3). It can be seen that $\sigma(\omega)$ has exactly the same variations as $\alpha(\omega)$, $k(\omega)$ and $\varepsilon_2(\omega)$, and indicates the conditioned absorption of the incoming energies. The larger absorption and photoconductivity in perovskite compounds based on α-, β-, and γ-phases of CsSnI$_{3-x}$Cl$_x$, observed respectively at x=1 and x = 0, may be due to the smaller magnitude of their band gap.

### 3.4. Elastic properties

Modeling of mechanical properties of perovskites of the α-CsSnI$_{3-x}$Cl$_x$ family, namely, determination of their elastic constants, allows us to obtain a detailed understanding of the material behavior under macroscopic pressure. From this point of view, in this work, first-principles calculations are carried out to obtain relaxed structures of crystals of the α-CsSnI$_{3-x}$Cl$_x$ family (x = 0, 1, 2, 3) under pressure in the range from 0 to 30 GPa, in which the external pressure induction increases in steps of 5 GPa. The obtained results will allow us to understand the nature of the investigated materials in terms of their natural response to pressure-like external perturbations.

Under the action of pressure, the structural parameters of the materials change markedly. The calculated lattice constant $a$ of α-CsSnI$_{3-x}$Cl$_x$ (x = 0, 1, 2, 3) are summarized in Table 4 and plotted in Figure 9. The results obtained show that with increasing pressure the crystal constants of all compounds decrease.

Table 4. Calculated values of lattice constants a of
α-CsSnI$_{3-x}$Cl$_x$ (x = 0, 1, 2, 3) within the GGA framework.

| Pressure (GPa) | 0 | 5 | 10 | 15 | 20 | 25 | 30 |
|---|---|---|---|---|---|---|---|
| System\parameter | | | | $a$ (Å) | | | |
| CsSnI$_3$ | 6.259 | 5.853 | 5.643 | 5.498 | 5.39 | 5.302 | 5.227 |
| CsSnI$_2$Cl | 6.304 | 5.909 | 5.699 | 5.556 | 5.445 | 5.356 | 5.282 |
| CsSnICl$_2$ | 6.291 | 5.902 | 5.693 | 5.548 | 5.437 | 5.347 | 5.27 |
| CsSnCl$_3$ | 5.617 | 5.333 | 5.165 | 5.044 | 4.949 | 4.871 | 4.804 |

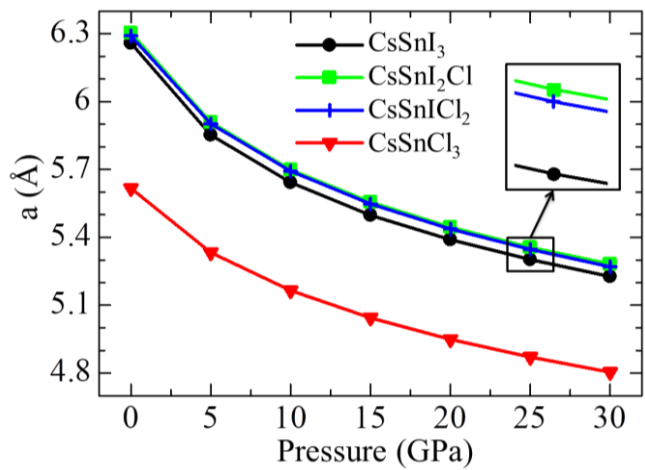

Figure 9. Calculated lattice constant a of α-CsSnI$_{3-x}$Cl$_x$ (x = 0, 1, 2, 3) as a function of pressure

The calculated pressure dependence of the relative volume change V/V$_0$ for α-CsSnI$_{3-x}$Cl$_x$ (x = 0, 1, 2, 3) is plotted in Figure 10, which is fitted to the Murnaghan's equation of state [69] expressed by:

$$V/V_0 = (1 + P\frac{B'}{B})^{-\frac{1}{B'}},\qquad(9)$$

where V and V$_0$ are volumes at pressure P and ambient pressure, respectively, and B and B' are bulk modulus and its pressure derivative, respectively. With increasing pressure, the relative volume of the lattice decreases. When moving from I to Cl, the relative volume of the systems under study increases.

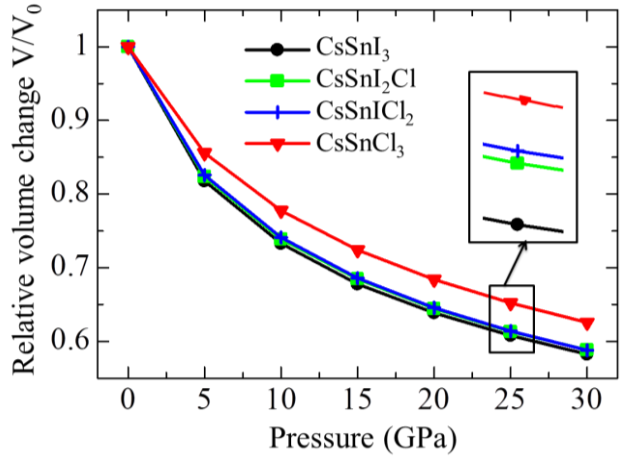

Figure 10. Calculated relative volume $V/V_0$ of $\alpha$-CsSnI3-xClx (x = 0, 1, 2, 3) as a function of pressure

The elasticity constant characterizes the stiffness of the crystal in response to external deformation, which can reflect the mechanical stability of the structure [70]. For this purpose, the elasticity coefficients $C_{ij}$ describing the mechanical and dynamic behavior of the crystal are used. The cubic crystal system $CsSnI_{3-x}Cl_x$ has three necessary constants, $C_{11}$, $C_{12}$ and $C_{44}$. For cubic symmetry, perovskite materials must satisfy the Born stability criteria: $C_{11} > 0$, $C_{44} > 0$, $C_{11}-C_{12} > 0$ and $C_{11} + 2C_{12} > 0$. The values we obtained satisfy the fundamental stability requirements [75], except for $CsSnI_2Cl$ and $CsSnICl_2$ at pressures above 20 Gpa. Knowing the elastic constants, we can calculate the values of bulk modulus (B) using Eq.:

$$B = \frac{C_{11} + 2C_{12}}{3}.$$ (10)

The elastic constants of the perovskite $\alpha$-CsSnI$_{3-x}$Cl$_x$ family are calculated using the stretch-strain method [71] implemented in VASP, and the bulk moduli B are summarized in Table 5 and presented in Figure 11 as a function of pressure.

Coefficient $C_{11}$ characterizes the shear pressure, which increases from 35.5 to 253.8, from 37.5 to 253.8, from 41.4 to 259.4, and from 47.9 to 244.7 for CsSnI$_3$, CsSnI$_2$Cl, CsSnICl$_2$, and CsSnCl$_3$, respectively, with increasing pressure from 0 to 30 Gpa. In turn, $C_{12}$ details the lateral growth of the materials, which also increases from 5.01 to 21.5, from 5.17 to 6.73, from 5.98 to 13.9, and from 7.9 to 25.2 for CsSnI$_3$, CsSnI$_2$Cl, CsSnICl$_2$, and CsSnCl$_3$, respectively, with increasing pressure. CsSnI$_3$ and CsSnCl$_3$ exhibit cubic stability up to 30 GPa as they satisfy the corresponding stability conditions. The compounds CsSnI$_2$Cl and CsSnICl$_2$ exhibit cubic stability up to 20 GPa, but beyond this pressure threshold, the elastic constant $C_{44}$ becomes negative, resulting in loss of cubic stability.

Table 5. Pressure dependence of the elastic constants $C_{ij}$ (all in GPa) and bulk moduli $B$ for the $\alpha$- CsSnI$_{3-x}$Cl$_x$ (x = 0, 1, 2, 3).

| System | Pressure | $C_{11}$ | $C_{12}$ | $C_{44}$ | $B$ |
|---|---|---|---|---|---|
| CsSnI$_3$ | 0 | 35.50114 | 5.01589 | 5.92882 | 15.17764 |
| | 5 | 81.17576 | 8.37204 | 7.26883 | 32.63995 |
| | 10 | 117.8536 | 9.0661 | 8.19573 | 45.32861 |
| | 15 | 156.116 | 13.12337 | 9.03331 | 60.78759 |
| | 20 | 188.5391 | 14.69204 | 9.85631 | 72.64104 |
| | 25 | 222.2462 | 18.54959 | 10.7723 | 86.44844 |
| | 30 | 253.8087 | 21.58532 | 11.44541 | 98.99311 |
| CsSnI$_2$Cl | 0 | 37.58026 | 5.17644 | 5.89161 | 15.97771 |

| | | | | |
|---|---|---|---|---|
| | 5 | 80.97833 | 5.09476 | 4.25839 | 30.38928 |
| | 10 | 119.374 | 4.75857 | 2.90674 | 42.96371 |
| | 15 | 155.4265 | 4.78352 | 1.38603 | 54.99783 |
| | 20 | 190.3706 | 5.43216 | -0.18479 | 67.07832 |
| | 25 | 223.0629 | 6.14079 | -1.79044 | 78.44816 |
| | 30 | 254.163 | 6.73404 | -3.47802 | 89.21037 |
| CsSnICl$_2$ | 0 | 41.49006 | 5.98859 | 6.03352 | 17.82241 |
| | 5 | 86.05495 | 7.58443 | 4.41492 | 33.74127 |
| | 10 | 125.7027 | 8.8823 | 2.85454 | 47.82242 |
| | 15 | 162.9949 | 10.63699 | 1.12682 | 61.42296 |
| | 20 | 197.5641 | 11.91841 | -0.76927 | 73.8003 |
| | 25 | 229.9778 | 13.44107 | -2.45473 | 85.61997 |
| | 30 | 259.425 | 13.9859 | -5.5291 | 95.64656 |
| CsSnCl$_3$ | 0 | 47.93337 | 7.99036 | 7.67982 | 21.3047 |
| | 5 | 88.18753 | 10.92013 | 7.45618 | 36.67593 |
| | 10 | 122.7583 | 13.42084 | 6.81187 | 49.86665 |
| | 15 | 155.9491 | 16.28496 | 5.94841 | 62.83966 |
| | 20 | 186.3445 | 18.90914 | 5.2206 | 74.72093 |
| | 25 | 216.1941 | 22.26685 | 4.64435 | 86.90926 |
| | 30 | 244.7027 | 25.26311 | 3.6989 | 98.40963 |

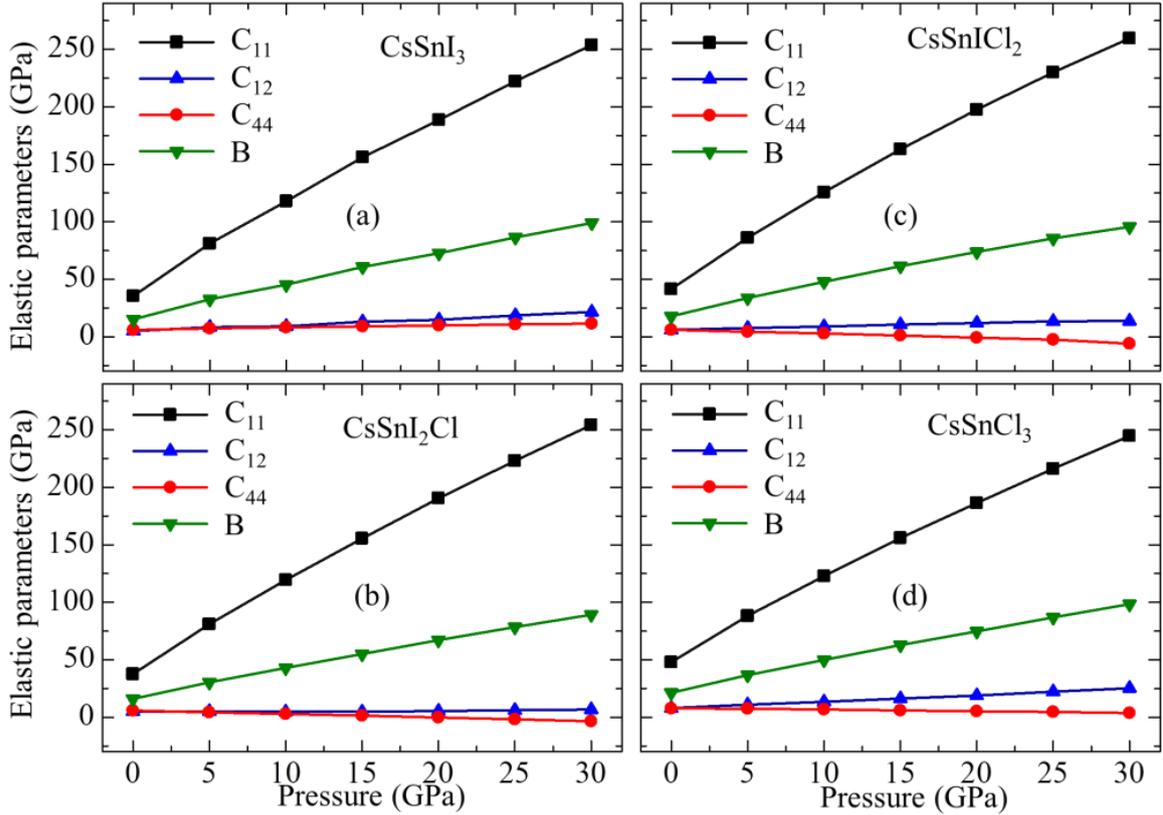

Figure 11. Calculated elastic constants C$_{ij}$ and bulk moduli $B$ for
α- CsSnI$_{3-x}$Cl$_x$ (x = 0, 1, 2, 3) as a function of pressure

Using these calculated elastic constants, sound velocities can be obtained. The mean sound velocity $v_m$ is expressed [73] in terms of the longitudinal sound velocity $v_l$ and the transverse one $v_t$ as:

$$v_m = \left[\frac{1}{3}\left(\frac{2}{v_t^3} + \frac{1}{v_l^3}\right)\right]^{-1/3},$$

(11)

in which $v_l$ and $v_t$ are calculated [74] by:

$$v_l = \sqrt{\frac{3B + 4G}{3\rho}} \quad \text{and} \quad v_t = \sqrt{\frac{G}{\rho}}, \tag{12}$$

where

$$G = \frac{(G_V + G_R)}{2} \begin{cases} G_V = \frac{(C_{11} - C_{12} + 3C_{44})}{5} \\ G_R = \frac{5(C_{11} - C_{12})C_{44}}{[4C_{44} + 3(C_{11} - C_{12})]} \end{cases}.$$

Here: $G$ is the isotropic shear modulus, $G_V$ is the Voight's shear modulus (an upper limit for G), and $G_R$ is the Reuss's shear modulus (a lower limit for $G$).

The Debye temperature, which is sensitive to various physical parameters such as melting point, phonon heat capacity and elastic constants, can be calculated using Eq. [74]:

$$\Theta_D = \frac{h}{k}\left(\frac{3n}{4\pi}\left(\frac{N_A \rho}{M}\right)\right)^{1/3} v_m, \tag{13}$$

where $h$ and $k$ are the Planck's and Boltzmann's constants, respectively. $N_A$ is the Avogadro's number, $\rho$ is the density, $M$ is the molecular weight, and $n$ denotes the number of atoms per formula unit (here five for α-CsSnI$_{3-x}$Cl$_x$).

Table 6. Calculated shear moduli, sound velocities and Debye temperatures for α-CsSnI$_{3-x}$Cl$_x$

| System | P, GPa | $G$, GPa | $v_t$, m/s | $v_l$, m/s | $v_m$, m/s | $\Theta_D$, K |
|---|---|---|---|---|---|---|
| CsSnI$_3$ | 0 | 8.750501 | 1429.506 | 2503.809 | 1588.557 | 140.5390 |
| | 5 | 14.80675 | 1681.55 | 3162.803 | 1878.963 | 177.7619 |
| | 10 | 19.54382 | 1828.867 | 3495.32 | 2045.809 | 200.7492 |
| | 15 | 23.9522 | 1947.117 | 3831.026 | 2182.152 | 219.7754 |
| | 20 | 27.97793 | 2042.693 | 4049.328 | 2290.312 | 235.2907 |
| | 25 | 31.98697 | 2130.877 | 4280.861 | 2391.072 | 249.7192 |
| | 30 | 35.60567 | 2200.648 | 4463.354 | 2470.712 | 261.7390 |
| CsSnI$_2$Cl | 0 | 8.959554 | 1580.841 | 2790.822 | 1757.899 | 146.5787 |
| | 5 | 12.16749 | 1671.831 | 3272.227 | 1873.028 | 166.6186 |
| | 10 | 14.67662 | 1739.129 | 3589.812 | 1954.447 | 180.268 |
| | 15 | 16.62113 | 1781.535 | 3838.472 | 2006.457 | 189.8283 |
| | 20 | 18.28421 | 1812.824 | 4054.401 | 2045.14 | 197.4324 |
| | 25 | 19.64644 | 1833.258 | 4230.947 | 2070.851 | 203.2365 |
| | 30 | 20.74578 | 1844.944 | 4378.975 | 2086.247 | 207.6159 |
| CsSnICl$_2$ | 0 | 9.459279 | 1776.437 | 3186.441 | 1977.966 | 155.3741 |
| | 5 | 12.5939 | 1862.602 | 3731.025 | 2089.684 | 174.969 |
| | 10 | 14.84213 | 1915.584 | 4088.507 | 2156.446 | 187.1876 |
| | 15 | 16.50368 | 1943.285 | 4369.196 | 2192.81 | 195.3189 |
| | 20 | 17.68917 | 1951.794 | 4579.607 | 2206.142 | 200.5181 |
| | 25 | 18.84025 | 1964.49 | 4762.768 | 2223.078 | 205.4585 |
| | 30 | 17.67226 | 1861.672 | 4835.181 | 2111.187 | 197.9683 |
| CsSnCl$_3$ | 0 | 11.39221 | 1842.915 | 3298.48 | 2051.627 | 167.3162 |
| | 5 | 15.46876 | 1986.688 | 3823.692 | 2223.384 | 190.9796 |
| | 10 | 18.21849 | 2054.974 | 4145.997 | 2306.468 | 204.5602 |
| | 15 | 20.44157 | 2100.702 | 4410.195 | 2362.887 | 214.5912 |
| | 20 | 22.48657 | 2141.324 | 4620.622 | 2411.843 | 223.2418 |
| | 25 | 24.53656 | 2184.131 | 4822.611 | 2462.653 | 231.595 |
| | 30 | 26.06829 | 2204.984 | 4983.655 | 2488.659 | 237.3047 |

The calculated sound velocities $v_m$, $v_t$ and $v_l$, shear modulus $G$ and Debye temperature $\Theta_D$ as a function of pressure, which are examples of the fundamental mechanical properties of α-CsSnI$_{3-x}$Cl$_x$ to determine their hardness or stability, are summarized in Table 6 and are plotted in Figures 12, 13a, and 13b for sound velocities, shear moduli, and Debye temperatures, respectively. The $v_m$, $v_t$, $v_l$, $G$, and $\Theta_D$ values for all systems increase with pressure.

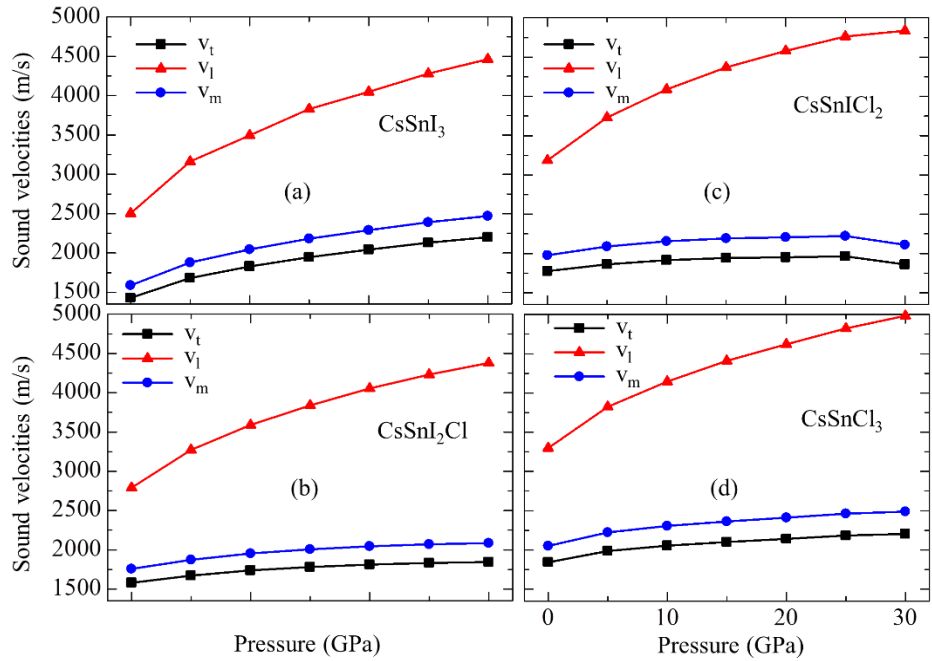

Figure 12. Calculated sound velocities for α-CsSnX$_3$ (X = I, Br and Cl) as a function of pressure

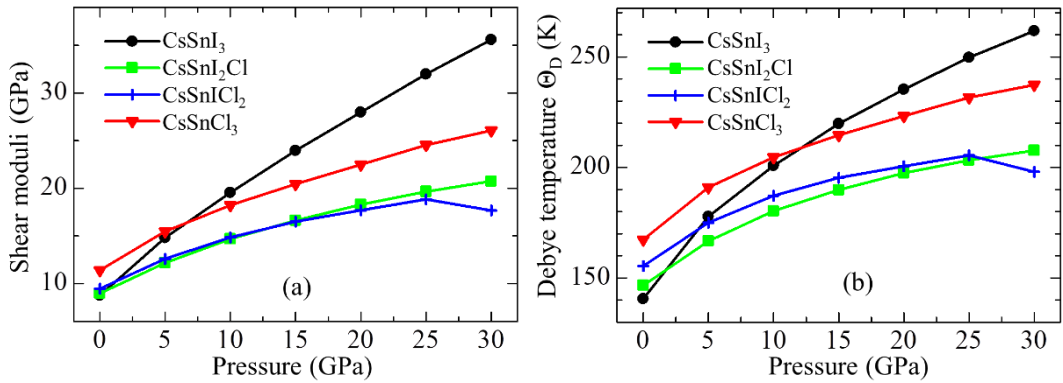

Figure 13. Calculated (a) shear moduli and (b) Debye temperatures
for α-CsSnI$_{3-x}$Cl$_x$ as a function of pressure

## 4. Conclusions

The structural, electronic, optical and elastic properties of compounds such as α-, β- and γ-phases CsSnI$_{3-x}$Cl$_x$ (x = 0, 1, 2, 3) were studied DFT using GGA, SCAN and HSE06. As a result, it is found that the position I(1) is energetically more favorable for Cl ions in CsSnI$_3$. We find that the lattice constants of CsSnI$_{3-x}$Cl$_x$ decrease with decreasing the size of anions from I to Cl. It is found that the band gap of CsSnI$_{3-x}$Cl$_x$ at transition from CsSnI$_3$ to CsSnCl$_3$ first slightly decreases and then increases. This trend appears to be the same in all investigated CsSnI$_{3-x}$Cl$_x$ phases. The total and partial densities of states reflecting the semiconducting properties of the studied systems have been calculated and discussed. The obtained lattice parameters and band gap of the studied systems are in good agreement with experimental and other theoretical works. Optical properties such as complex dielectric function, absorption coefficient, reflectivity, refractive index, extinction coefficient, optical conductivity and energy loss function for pure and mixed cases are discussed. The low reflectivity, wide absorption range and high absorption coefficients of these compounds make them good candidates for various optoelectronic and photovoltaic applications.

The mechanical properties of α-CsSnI$_{3-x}$Cl$_x$ have been investigated using the GGA approximation. The elastic constants C$_{ij}$ for α-CsSnI$_{3-x}$Cl$_x$ were calculated and their bulk moduli, shear moduli, sound velocities and Debye temperatures were discussed. The materials are found to be mechanically stable perovskites according to Born stability criteria and low bulk moduli. Thus, in general, perovskites of the CsSnI$_{3-x}$Cl$_x$ family are good candidates for optical materials used in solar cells. The results obtained in this work provide a theoretical basis for the experimental study of mixed perovskites, allowing further investigation of other types of materials with suitable properties and finding potential applications in the semiconductor and optoelectronic industries.

## Funding

The work was performed at the S.U. Umarov Physical-Technical Institute of the National Academy of Sciences of Tajikistan with the support of International Science and Technology Center (ISTC), project TJ-2726.